\documentclass[%
preprint,
superscriptaddress,
 amsmath,amssymb,
 aps,longbibliography,
floatfix,
]{revtex4-1}

\usepackage{graphicx}%
\usepackage{dcolumn}%
\usepackage{bm}%
\usepackage{verbatim}
\usepackage[version=4]{mhchem}
\usepackage{xcolor}
\usepackage{physics}
\usepackage[normalem]{ulem}
\usepackage{pdfpages}
\usepackage{pgffor}
\usepackage{makecell}
\usepackage{grffile}
\makeatletter
\AtBeginDocument{\let\LS@rot\@undefined}
\makeatother

\begin{document}

\preprint{APS/123-QED}

\title{Ranking the Synthesizability of Hypothetical Zeolites with the Sorting Hat}

\author{Benjamin A.\ Helfrecht}
 \affiliation{Laboratory of Computational Science and Modeling, Institut des Mat\'eriaux, \'Ecole Polytechnique F\'ed\'erale de Lausanne, 1015 Lausanne, Switzerland}

\author{Giovanni Pireddu}
\affiliation{PASTEUR, D\'epartement de Chimie, \'Ecole Normale Sup\'erieure, PSL University, Sorbonne Universit\'e, CNRS, 24 rue Lhomond, 75005 Paris, France}

\author{Rocio Semino}
\email{rocio.semino@umontpellier.fr}
\affiliation{ICGM, Univ. Montpellier, CNRS, ENSCM, Montpellier, France}

\author{Scott M.\ Auerbach}
\email{auerbach@umass.edu}
\affiliation{
 Department of Chemistry and Department of Chemical Engineering, University of Massachusetts Amherst, Amherst, MA 01003 USA\\
}

\author{Michele Ceriotti}
\email{michele.ceriotti@epfl.ch}
 \affiliation{Laboratory of Computational Science and Modeling, Institut des Mat\'eriaux, \'Ecole Polytechnique F\'ed\'erale de Lausanne, 1015 Lausanne, Switzerland}

\date{\today}%

\begin{abstract}
Zeolites are nanoporous alumino-silicate frameworks widely used as catalysts and adsorbents. Even though millions of distinct siliceous networks can be generated by computer-aided searches, no new hypothetical framework has yet been  synthesized. The needle-in-a-haystack problem of finding promising candidates among large databases of predicted structures has intrigued materials scientists for decades; most work to date on the zeolite problem has been limited to intuitive structural descriptors. Here, we tackle this problem through a rigorous data science  scheme---the “zeolite sorting hat”---that exploits interatomic correlations to produce a 95\% real versus theoretical zeolites classification accuracy.
The hypothetical frameworks that are grouped together with known zeolites are promising candidates for synthesis, that can be further ranked by estimating their thermodynamic stability. A critical analysis of the classifier reveals the decisive structural features. Further partitioning into compositional classes provides guidance in the design of synthetic strategies.
\end{abstract}

\maketitle

\section{Introduction}

Zeolites are nanoporous crystalline materials with exceptionally high thermal and hydrothermal stabilities, making them excellent candidates for a range of present and future technologies based on shape selectivity. Because of their controlled nanoporosity and acidic properties, zeolites find application in myriad industrially-relevant processes, predominantly in separation and catalysis.\cite{ZeoHandbook2003} 
To accelerate zeolite discovery, databases of hypothetical zeolites have been created \cite{Friedrichs1999,Li2003,Treacy2004,Pophale2011} containing millions of new framework structures. 
Even though these databases have been successfully screened identifying materials with desirable properties\cite{lin+02nm,evan-coud17cm,moli+19acr,jabl+20cr}, but to date, none of them has been synthesized in the lab, a phenomenon referred to as the ``zeolite conundrum.'' \cite{Blatov2013} 
The great importance and challenge in fabricating new zeolites prompt several pressing questions: How do collections of real and hypothetical zeolites relate to each other in terms of structural diversity? Which structural features play the biggest role in distinguishing real and hypothetical zeolites? Which hypothetical zeolites are most likely synthesizable, and in which chemical composition? 
Previous attempts to answer these questions 
\cite{Majda2008,Dawson2012,Li2013,Liu2015,Lu2017,Salcedo2019,Kuznetsova2018,Li2019} have relied on intuitive guesses for structural descriptors such as rings and angles, which provide incomplete \cite{Helfrecht2019} and thus biased results.
In the present work, we answer all these questions via rigorous data science methods combining unsupervised and supervised machine learning,\cite{Ceriotti2019} along with the generalized convex hull (GCH) description of thermodynamic stability,\cite{Anelli2018} yielding a new and powerful approach for sorting real\cite{ZeoAtlas2007} and hypothetical\cite{Friedrichs1999,Li2003,Treacy2004,Pophale2011} zeolites, as well as finding promising zeolite candidates and suggesting likely chemical compositions for them.

\begin{figure*}
    \centering
    \includegraphics[width=\textwidth]{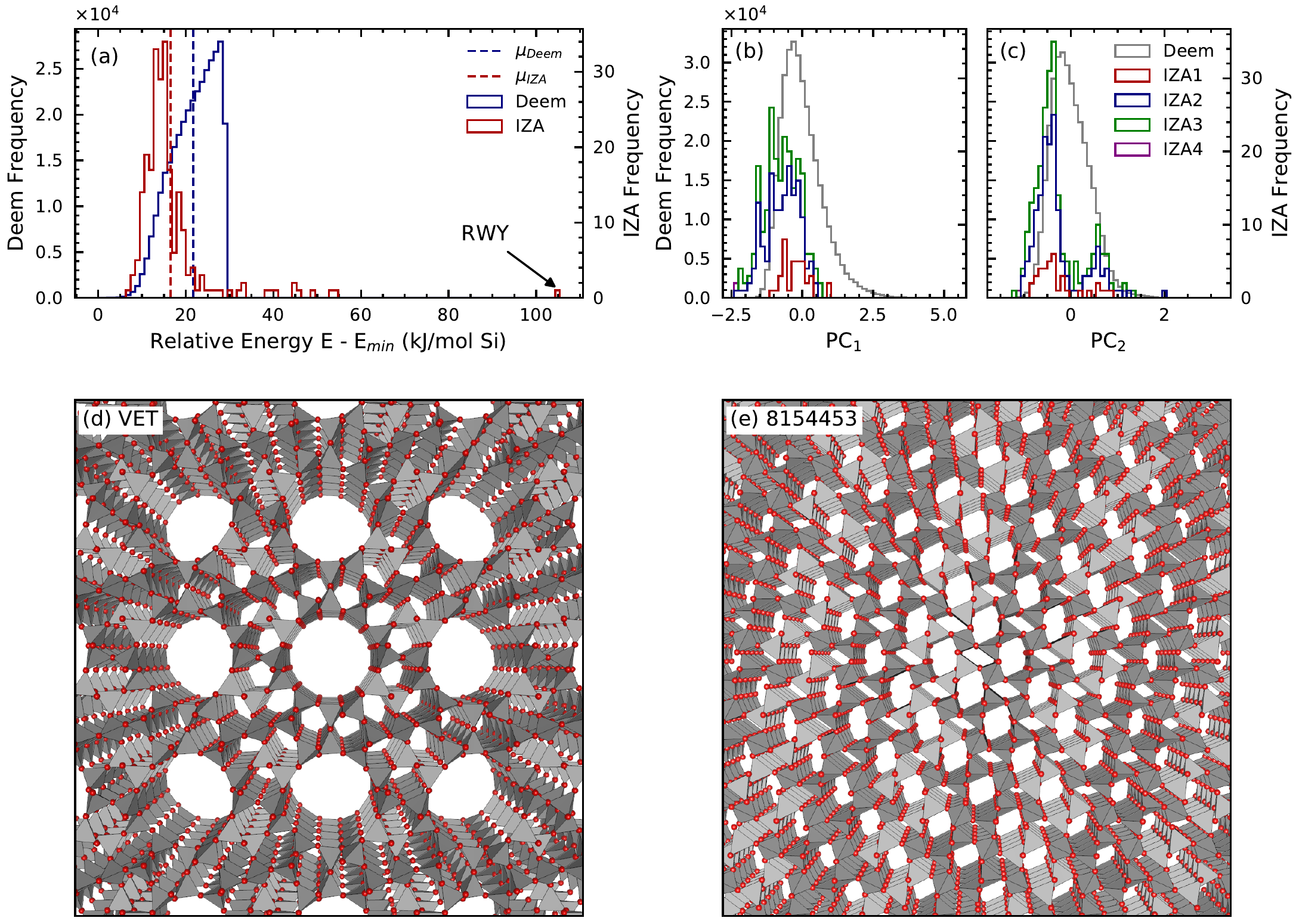}
    \caption{
        Histograms of (a) energies computed for the IZA and DEEM frameworks with GULP
        and (b)-(c) of values of the first two principal components
        of the power spectrum SOAP vectors
        of a subset of 10,000 DEEM frameworks and all 230 IZA frameworks.
        The histogram makes evident that the IZA frameworks are concentrated
        near the edge of the structural space defined by the DEEM frameworks.
        The PCA projection is defined only by the 10,000 DEEM frameworks.
        (d) Atomic snapshot of PON, the IZA framework based with the lattice energy closest to the IZA average.
        (e) Atomic snapshot of framework 8183215, the DEEM structure with the lattice energy closest to the DEEM average.
    }
    \label{fig:pca_histogram}
\end{figure*}

\section{Results} 

The scale of the zeolite conundrum can be appreciated by comparing the number of  hypothetical frameworks with that of ``real'' zeolites. 
Different studies have suggested over 2,600,000 distinct topologies \cite{Pophale2011}, and even the subset we consider here, which only contains particularly stable, fully connected frameworks selected among a larger pool of candidates, contains more than 300,000 all-silica structures (Ref.~\citenum{Pophale2011}, by Deem and coworkers, henceforth denoted as DEEM). 
In contrast, only 255 framework topologies have been collected in the International Zeolite Association database (henceforth denoted IZA), which can be realized in different compositional variations. To ensure that our comparisons are made on an equal footing, we perform our study on all-silica models.
The great imbalance between known and hypothetical frameworks calls for a balancing act when applying data-driven analyses: models and structural descriptors must be flexible and sensitive enough to detect structural differences among all of the DEEM frameworks, but sufficiently robust and concise to extract useful information from a few hundred IZA entries without overfitting the smaller dataset. 

To this end, we describe framework structures using the Smooth Overlap of Atomic Postions (SOAP) method \cite{bart+13prb}, which allows systematic convergence of structural information by increasing the SOAP length scale and the order of atomic correlations (distances, angles, dihedrals, … \cite{will+19jcp,musi+21cr}). In previous work, we proved this convergence by applying SOAP to machine-learn framework density and lattice energy of DEEM frameworks \cite{Helfrecht2019}. The heuristic fingerprints that have been used previously to distinguish real and hypothetical zeolites do not allow such convergence, and thus are almost certainly incomplete. Here we find that the DEEM-trained models accurately predict the same properties for IZA frameworks (see Table S1, energy error below 0.20 kJ/mol-Si). The accuracy of DEEM-trained predictions on IZA indicates substantial structural overlap between the two datasets – underscoring the significant challenge of telling them apart.

Armed with SOAP as a structural descriptor, we seek a method for discriminating IZA and DEEM entries. We note that classifiers based on lattice energy (Fig.\ \ref{fig:pca_histogram}(a)) or unsupervised learning (Figs.\ \ref{fig:pca_histogram}(b) and \ref{fig:pca_histogram}(c)) are both found to fail at telling IZA and DEEM apart (also see point-clouds in Fig.\ S1). To solve this puzzle, we apply {\it supervised} learning to classifying zeolites, and denote our approach the “Zeolite Sorting Hat.”

We actually seek to solve an even harder problem -- distinguishing subclasses of IZA based on composition. Applying the same criterion to DEEM frameworks suggests the chemical composition that should be pursued in the laboratory for a candidate structure -- making our predictions of synthesizability more directly useful for materials chemists. 
To do this, we parse IZA into subclasses (or ``houses'') based on reference compositions -- i.e., the chemical composition of the first entry listed in the IZA database for each topology. Our parsing method follows from the premise that the composition of the first instance of a given zeolite topology points to a convenient synthetic approach, that can also be applied to realize a structurally similar DEEM framework. 
This yields the following four IZA houses: zeolite topologies with a pure-silica reference composition are assigned to IZA1; topologies whose reference composition contains O, but no Si, are classified as IZA3; topologies referenced to an intermediate fraction of Si (e.g., aluminosilicates) are labeled as IZA2; a single exotic framework (RWY) containing neither Si nor O is classified as IZA4 and is discarded from the present analysis as a structural and energetic outlier.
Through the lens of the principal SOAP components shown in Figs.\ \ref{fig:pca_histogram}(b,c), these IZA houses occupy the same region in SOAP vector space, thus appearing indistinguishable according to these two principal component directions.
In summary then, the various framework classes (DEEM/IZA1/IZA2/IZA3) cannot be effectively discriminated using an energetic criterion or through unsupervised learning. %

We have designed the Zeolite Sorting Hat to give both accuracy and interpretability. The Zeolite Sorting Hat involves a linear support vector machine (SVM) whose inputs are the SOAP vectors and whose outputs are decision functions: one function for the two-class sort (DEEM/IZA) and four functions for the four-class sort (DEEM/IZA1/IZA2/IZA3). 
We use the magnitude of the decision functions to provide a more nuanced assessment of the classification, allowing us to rank the ``IZA-ness'' of a set of DEEM structures. 
The Zeolite Sorting Hat was trained on a random half of IZA, and approximately 3\%{} of DEEM (see Methods and the SI for details), and the results we report here refer to predictions made on the remainder of the datasets.  

\begin{figure*}
    \centering
    \includegraphics[width=0.8\textwidth]{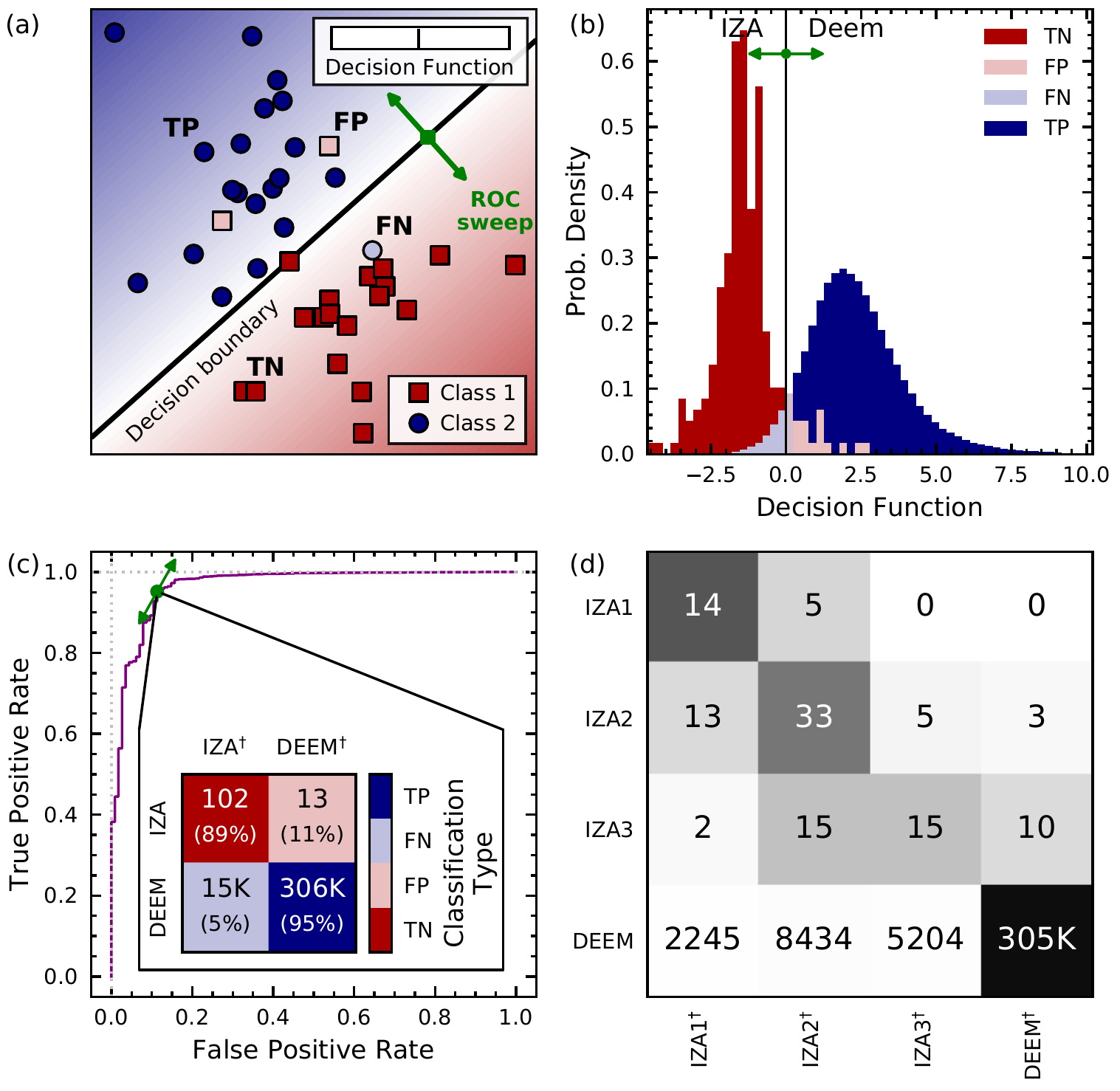}
    \caption{
        (a) Schematic of a support vector machine (SVM); each dot or square represents the feature vector for a given data point, the shading represents the value of the decision function, and the decision boundary location can be adjusted to optimize the classification (represented by a green arrow)
        (b) Histogram of decision-function values for IZA and DEEM frameworks
        based on the SOAP power spectrum with an environment cutoff of 6.0 \AA{};
        (c) Receiver operating characteristics (ROC) curve for the IZA vs.\ DEEM SVM classification with 6.0 \AA{}
        SOAP, as the decision-function boundary is swept through
        decision space as shown by green arrows in (a), (b), and (c). The inset in (c)
        shows the confusion matrix for the two-class IZA vs. DEEM classification using
        the full SOAP power spectrum, and (d) similarly shows the four-class confusion matrix,
        with darker shading indicating a greater proportion of the class-wise predictions.
        The superscripts $^{\dagger}$ in
        confusion matrix labels refer to predicted classifications, and the labels
        TP, FP, TN, and FN indicate true positive, false positive, true negative,
        and false negative classifications, where the DEEM and IZA frameworks 
        are denoted as the positive and negative classes, respectively.
    }
    \label{fig:svc_df_roc}
\end{figure*}

Figure \ref{fig:svc_df_roc} shows how the Zeolite Sorting Hat works and displays its performance in classifying real (IZA, red) and hypothetical (DEEM, blue) frameworks. 
Figure \ref{fig:svc_df_roc}(b) reveals the actual histogram of decision-function values for the two-class IZA/DEEM sort obtained from the full SOAP power spectrum including two- and three-body correlations within a distance cutoff of 6.0 \AA{}. The histogram is clearly bimodal, indicating that the IZA and DEEM datasets are indeed distinguishable via the Zeolite Sorting Hat, a striking contrast to the failure of unsupervised learning shown in Fig.\ \ref{fig:pca_histogram}. 
Figure \ref{fig:svc_df_roc}(c) quantifies the performance of the Zeolite Sorting Hat through the receiver operating characteristic (ROC) curve and optimal confusion matrix. The ROC curve optimizes sorting accuracy with respect to the location of the decision boundary (green arrows in Fig.\ \ref{fig:svc_df_roc}) by maximizing the rate of true positives while minimizing false positives. 
The best Zeolite Sorting Hat performance is shown by the confusion matrix inset in Fig.\ \ref{fig:svc_df_roc}(c), revealing that 102/115 (89\%) of the IZA frameworks and 95\% of DEEM frameworks are correctly classified. This excellent sorting performance is remarkable given the substantial overlap in energy- and structure-spaces shown in Fig.\ \ref{fig:pca_histogram}.

The successful two-class (DEEM/IZA) sort prompts the investigation of the even more challenging four-fold (DEEM/IZA1/IZA2/IZA3) classification: a prediction of the most easily synthesizable composition based exclusively on the structure of the pure \ce{SiO2} framework. 
The optimal confusion matrix of the four-way classifier (Fig.\ \ref{fig:svc_df_roc}(d)) demonstrates that the Zeolite Sorting Hat is also successful at this more difficult task. 
The distinction between IZA1 (all-silica) and IZA3 (no-silicon) is nearly perfect. Most of the incorrect classifications involve IZA2, a house that contains a broad range of compositions from high-silica alumino-silicates to low-silicon silico-aluminophosphates, justifying the overlaps with IZA1 and IZA3 houses. 
It is intriguing to note that most of the IZA being misclassified as DEEM belong to the no-silicon IZA3 house, while the all-\ce{SiO2} IZA1 entries are never mistaken for a hypothetical framework.
The success of this four-fold sort opens the door to recommending synthesis compositions for IZA-like DEEM structures.

The confusion matrix in Fig.\ \ref{fig:svc_df_roc}(c) shows that $\approx$15,000 DEEM frameworks are misclassified as IZA. To further narrow down the subset of DEEM structures for which synthesis should be attempted, we augment the notion of similarity generated by the Zeolite Sorting Hat with the concept of thermodynamic stability provided by the convex hull \cite{ogan+19nrm}. %
The importance of thermodynamic stability was underscored by a recent machine-learning study showing that synthesizable zeolite phases correlate with their thermodynamic stabilities \cite{ma2020-chemsci}. 
However, as discussed above, Fig.\ \ref{fig:pca_histogram}(a) shows that naively using lattice energies to identify synthesizable DEEM frameworks is insufficient because of the significant overlap between IZA and DEEM energetics. 
Furthermore, most zeolites are only metastable, their synthesis being made possible by carefully-chosen thermodynamic conditions that cannot be mapped onto a single stabilizing order parameter. 
For this reason, we have used a generalized convex hull (GCH) construction, which uses data-driven coordinates as proxies for composition and thermodynamic  variables \cite{anel+18prm}.

The Zeolite Sorting Hat supplies a natural, data-driven coordinate space for the GCH through the method of principal covariates regression (PCovR) \cite{dejo-kier92cils,helf+20mlst} -- optimizing a lower-dimensional space that supports classification by the Zeolite Sorting Hat. We have applied PCovR to produce a 2D latent space for the GCH. The energetics for the GCH were taken from the same classical forcefield used by Deem and coworkers\cite{SLC1984}. After determining which frameworks define the vertices of the GCH, the ``hull energy'' of each framework was determined by the vertical energy difference of the framework energy to the GCH. %

\begin{figure*}[bthp]
    \centering
    \includegraphics[width=\linewidth]{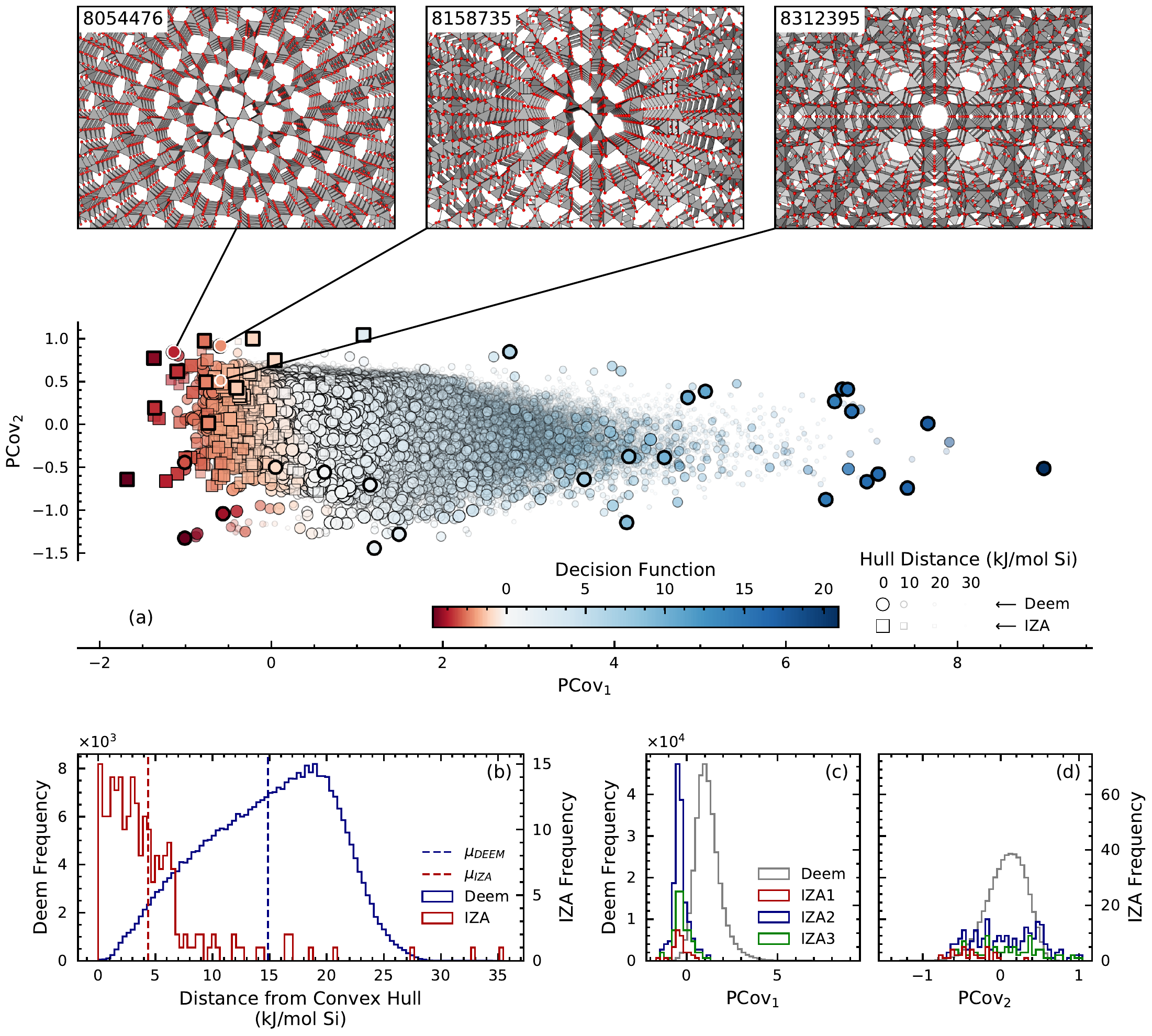}
    \caption{
        (a) First two components of the PCovR projection based on the four-class 
        decision functions with IZA (square) and DEEM (circle) frameworks colored
        according to two-class decision-function value (IZA-like = red; DEEM-like = blue). 
        Larger, more opaque points lie close in energy to the GCH; hull vertices
        indicated via points with thick black borders;
        (b) Histogram of the energy distance to the convex hull
        for the IZA and DEEM frameworks.
        (c)--(d) Histograms of the PCovR component values
        for the IZA houses and DEEM.
    }
    \label{fig:gch}
\end{figure*}

A visualization of the resulting GCH construction is given
in Fig.\ \ref{fig:gch}(a), in which each IZA and DEEM framework
is plotted as a single square or circle, respectively, and is colored according to its two-class DEEM/IZA decision-function value. The points are also sized
and given an opacity corresponding their hull energies: larger, more opaque points are those
closer to the GCH. The hull vertices are indicated via thick black borders. 
The DEEM frameworks that show the most promise according to our criteria, then, are those that are misclassified as IZA
and lie close to the GCH -- i.e., the large, opaque, red circles in Fig.\ \ref{fig:gch}(a).

The utility of the GCH construction based on a PCovR latent space 
is evident upon comparing Figs.\ \ref{fig:gch}(b)-\ref{fig:gch}(d) with
Figs.\ \ref{fig:pca_histogram}(a)-\ref{fig:pca_histogram}(c).
While the difference in mean IZA and DEEM lattice energies in Fig.\ \ref{fig:pca_histogram}(a) is only 5 kJ/mol-Si, the mean hull energies shown in Fig.\ \ref{fig:gch}(b) differ by more than 10 kJ/mol-Si with most IZA frameworks either on or very close to the GCH, thus confirming that the energy distance from the convex hull is a more selective filter for thermodynamic stability than the bare lattice energy.
Moreover, Figs.\ \ref{fig:gch}(c-d) reveal the advantage 
of the PCovR space over the PCA space 
(Figs.\ \ref{fig:pca_histogram}(b-c)) in arranging
the IZA and DEEM frameworks: the first PCovR component
correlates strongly with DEEM/IZA decision-function values,
and therefore highlights the structural distinction between the
real and hypothetical frameworks -- 
an arrangement that is largely absent from the PCA.
Furthermore, the second PCovR component (Fig.\ \ref{fig:gch}(d))
roughly organizes the IZA frameworks according to their
compositional (house) classification: there is minimal overlap between
the all-silica (IZA1) and no-silicon (IZA3) frameworks,
and the frameworks that contain both Si and other tetrahedral (``T'') atoms (IZA2) overlap with both the all-silica and no-silicon frameworks. Note that the separation into IZA houses through the Zeolite Sorting Hat shown by the confusion matrix (Fig.\ \ref{fig:svc_df_roc}(d)) is actually better than what can be visually inferred from the histogram in Fig.\ \ref{fig:gch}(d) because of the dimensionality reduction applied by the PCovR method. 

The GCH construction leaves us with approximately 4,700 DEEM structures that are classified by the Zeolite Sorting Hat as belonging to IZA and that lie within a 5 kJ/mol window from the hull. 
From energetic and structural perspectives, these frameworks appear as likely to be synthesizable as structures that have been made already.
While selecting among these worthwhile candidates for synthesis can be achieved by ranking the structures based on their GCH distances, a more application-oriented selection can be performed by introducing a secondary filtering criterion.
To demonstrate this approach, we stratify the DEEM dataset in terms of molar volume and select the IZA-like DEEM candidate closest in energy to the hull within each of the 55--60, 60--65, and 65--70 \AA$^3$/Si ranges, which cover the upper end of the distribution of molar volumes for known IZA structures. 

The structures of the resulting three promising DEEM frameworks are highlighted in the insets of Fig.\ \ref{fig:gch}(a). 
Two (8158735 and 8054476) are classified by the Zeolite Sorting Hat as belonging to IZA2, e.g., as aluminosilicates, and the third (8312395) is classified as IZA3, a zeolite containing no silicon -- suggesting synthesis as an aluminophosphate.
Framework 8158735, the candidate within the 55-60 \AA $^3$/Si range, is closest to the IZA framework THO in SOAP space and exhibits rings of 3, 4 and 8 T atoms; THO is similarly composed of 4- and 8-rings.
The candidate within the 60-65 \AA$^3$/Si range is the triclinic framework 8054476, having SBN as its closest IZA neighbor. Like SBN, framework 8054476 contains 4-, 8- and 9-member rings.
The DEEM framework in the largest volume category that is closest in energy to the GCH is structure 8312395, which shares many structural similarities to its nearest IZA neighbor RHO: both frameworks contain 4-, 6-, and 8-member rings. Additional discussion of the similarities of these three candidate DEEM frameworks and their nearest IZA analogues is given in the SI.

\begin{figure*}[tbhp]
\centering
\includegraphics[width=\textwidth]{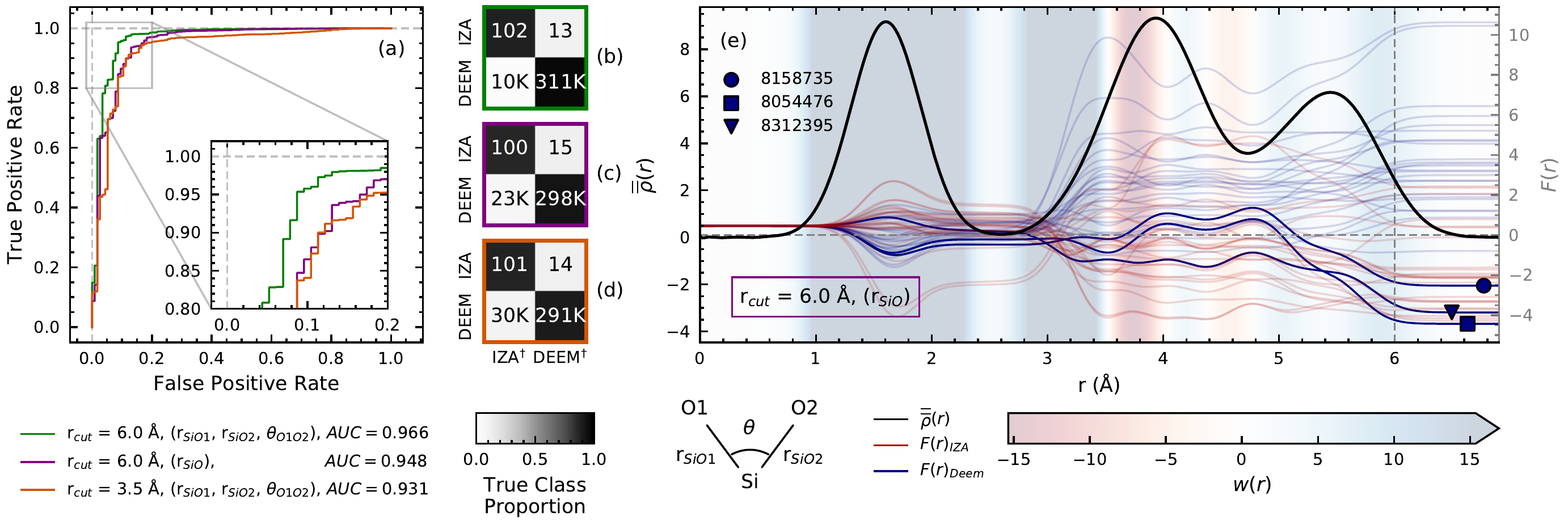}
\caption{
    (a) ROC curves of SVM classifications based on three ``knock-out'' models based 
    on a limited set of structural correlations between a Si and neighboring O atoms. 
    The models include angular and radial correlations up to 6.0 \AA{} (green), 
    only radial information (purple) and radial and angular correlations limited to 3.5 \AA{} (orange).
    The corresponding confusion matrices are shown in panels (b)--(d).
    (e) Class-averaged IZA and DEEM real-space densities based on Si-O correlations 
    (black line) plotted together with the decision traces $d(r)$ 
    for 25 DEEM frameworks (faded blue lines) and 25 IZA frameworks (faded red lines).
    The $d(r)$ for the three highlighted frameworks from Fig.\ \ref{fig:gch} 
    are also plotted as fully opaque lines and are labeled using symbols. 
    The line $F(r) = 0$ indicates the decision boundary: 
    the top half corresponds to DEEM predictions, the bottom half to IZA.
    The background coloring representing the SVM weights $w(r)$ is subject to a threshold
    to better show sign changes: weights falling outside the colorbar limits are assigned
    the color at the corresponding end of the colorbar.
    }
    \label{fig:svm_brain}
\end{figure*}

The Zeolite Sorting Hat's ability to discriminate between IZA and DEEM, and among different reference compositions of IZA, represents a breakthrough that begs a fundamental question: what aspects of zeolite structure are critical to these discriminating powers? We note that while previous studies of real and hypothetical zeolites have postulated such structural discriminants as inputs, these arise naturally as outputs from analyzing our models.

To reveal key structural discriminants, we first
performed an \emph{ablation study} in which we built several ``knock-out models'' that use a subset of the structural features. We repeat the two-class DEEM/IZA sort to determine the impact on classification performance of ($i$) restricting the range of correlations to first neighbors (up to 3.5\AA{}), ($ii$) considering only radial information on pair correlations, and ($iii$) using only some of the three-body angular correlations between Si and O neighbors. 
We show three representative models in Figs.\ \ref{fig:svm_brain}(a-d), and report a more systematic investigation in the SI. 

Restricting the range of correlations or discarding \emph{all} angular information leads to degradation of classification performance, indicating that the structural features that distinguish real and hypothetical zeolites involve angular correlations and patterns in the relative positions of second and third neighbor atoms -- i.e., at length scales beyond the typical indicators that have been hypothesized in previous studies \cite{Majda2008,Dawson2012,Li2013,Liu2015,Lu2017,Salcedo2019,Kuznetsova2018,Li2019}.

Second, our use of linear constructs -- SOAP vectors and linear support vector machines -- allows us to recast the Sorting Hat in a ``real space" form, to elucidate the spatial weights that discriminate real and hypothetical zeolites. 
The decision functions are then obtained by summing the values of these weight functions over all pairs and triplets of atoms in a structure. 
For a purely radial model, the decision process can then be interpreted as the incremental construction of a ``decision trace'' $d(r)$ (see Methods) that, for ${r\rightarrow\infty}$, gives the value used for classification. The length scales at which $d(r)$ undergoes large changes are those that control the classification.

Figure \ref{fig:svm_brain}(e) shows that, for most DEEM frameworks, $d(r)$ settles to plateau positive values around $r=3.5$\AA{}, corresponding to the onset of second-neighbor Si-O correlations (3-4\AA{}). 
Known IZA frameworks show an opposite behavior, drifting towards negative decision values in the same region. The role played by these second-neighbor Si-O correlations manifests itself in the sharp change of $w(r)$ from positive to negative values at around 3.5\AA,
indicating that frameworks in which the second-neighbor Si-O peak appears at shorter-than-average distances favor a DEEM prediction, and vice versa for IZA.

We thus conclude that second-neighbor Si-O distances are the most clearly discernable structural feature that differentiate IZA and DEEM frameworks.
More subtle structural correlations involving angular information and third-neighbor distances are needed to achieve classification accuracies above 90\%{}, making an automatic data analysis preferable to \textit{ad hoc} heuristics. 

In conclusion, synthesizing new zeolites is both an exciting intellectual challenge, and a technological quest with high potential rewards.  
The huge databases of computationally-proposed zeolites stand in striking contrast to the 255 known framework topologies (IZA), configuring a highly asymmetric problem that hinders the solution of this zeolite conundrum by brute-force applications of data science. 
We developed and applied a multi-pronged strategy for sifting through the hypothetical zeolites in search of the most promising candidates for synthesis. 
This ``Zeolite Sorting Hat'' tackles data scarcity by using flexible and unbiased SOAP structural descriptors as inputs, and relatively simple and robust linear classification algorithms via support vector machines to reach a 95\%{} accuracy in distinguishing real and hypothetical structures.  
The 5\%{} of hypothetical structures that are recognized as ``real'' by the Zeolite Sorting Hat become promising candidates for synthesis. A thermodynamic stability criterion provides an additional filter, and  together with stratification by framework density leads us to propose three leading hypothetical candidates for synthesis.
By further partitioning IZA frameworks into ``zeolite houses'' based on known reference compositions, and by quantifying geometric proximity to existing materials, we provide unique guidance for synthetic efforts at fabricating new zeolites. The principled choices we made in the architecture of the Zeolite Sorting Hat also allows to achieve a degree of interpretability in the classification process, pointing to the importance of second-neighbor Si-O distances as the leading factor that distinguishes real and hypothetical frameworks. 

As it is the cases for many synthetic tasks, making zeolites is a form of art, guided by experience, chemical intuition and serendipity. The Zeolite Sorting Hat introduces data-driven techniques and rationale design\cite{kuma+18jacs,rime18nc} into the process of selecting candidates that we hope will accelerate the rate of discovery, which in turn will improve the predictive capabilities of the model in a positive feedback mechanism that will progressively take the guesswork out of zeolite synthesis.

\section{Methods}

Zeolite structures were obtained as cif files from the IZA (\url{http://www.iza-structure.org/databases/}) and the hypothetical zeolites websites (\url{http://www.hypotheticalzeolites.net/DATABASE/DEEM/DEEM_PCOD/index.php}). 
IZA zeolites energies were calculated with GULP\cite{Gale2003} following the procedure previously considered for the DEEM zeolites in \citet{Pophale2011} with a modified version of the  
Sanders-Leslie-Catlow (SLC) potential\cite{SLC1984} to overcome the negative energy divergence due to the Buckingham contributions for $r\rightarrow0$ (see SI). Since DEEM frameworks are already relaxed with the SLC forcefield, we computed their energies though a relaxation of the atom shells only, keeping the cell and cores fixed. Our computed energies were in good agreement with those obtained by Deem and coworkers, except for five frameworks which were thus discarded from all of our analyses. We have also discarded DEEM frameworks that we found to be identical to an IZA framework. To establish this, we evaluated the Euclidean distance between 
the full power spectrum SOAP feature vectors with an environment cutoff of 6.0 \AA{} (more details can be found in the SI). 
SOAP representations were computed using \texttt{librascal}\cite{LIBRASCAL}
for two different environment cutoffs: 3.5 \AA{} and 6.0 \AA{}
for each Si-centered environment 
(that is, for every Si atom in a given framework). 
For our structure-based analyses, 
we define the SOAP representation of a given framework as the average 
over the SOAP vectors corresponding to each of its Si atoms. Because SOAP vectors have a high dimension, we applied principal component analysis (PCA) to find the dimensions that best describe the variance of the data. Further details of the SOAP calculations can be found in the SI.

Energies, molar volumes and compositions were predicted by optimizing the mean absolute error of the target property predictions via linear regression. 
Two-class (DEEM/IZA) and four-class (DEEM/IZA1/IZA2/IZA3) lineal kernel SVM models were built using sklearn to distinguish the IZA frameworks 
from the DEEM. SVM models were constructed for each combination of: 
SOAP environment cutoff (3.5 \AA{} or 6.0 \AA{}), 
n-body correlations (two-body radial spectrum and three-body power spectrum), 
and atom-atom correlations (Si-Si correlations, Si-O correlations, 
and Si-Si and Si-O correlations for the radial spectrum; 
and Si-Si-Si correlations, Si-O-Si correlations, and Si-O-O correlations 
and all combinations thereof for the power spectrum). The real-space expansion of the SOAP vectors was performed 
through a sum over the product of “contracted” Legendre DVR radial basis functions 
and spherical harmonics based on Legendre polynomials (for the power spectrum) 
with the SOAP vectors, summed over the expansion orders and angular index. The optimization target of the SVM was the class-balanced accuracy. To lend support to the validity 
of the class distinctions learned by the SVM, we assigned a set of random labels to the DEEM frameworks and subsequently attempted to classify them based on the randomly assigned labels. SVM was unable to learn the random labels, 
suggesting that the learned distinctions between the IZA and DEEM frameworks 
are indeed due to genuine differences in their structural characteristics. 
The ``decision trace" $d(r)$ is defined combining the smoothed radial correlation function $\rho(r)$ (the real-space counterpart of the pair descriptor associated with each structure) and a weight $w(r)$ that is also a function of distance, yielding
\begin{equation}
\label{eqn:cumulative_df}
    d(r) = b + \int_0^r \dd r^{\prime} [\rho(r^{\prime}) - \overline{\overline{\rho}}(r^{\prime})]  w(r^{\prime}).
\end{equation}
$d(r)$ is defined to also include the SVM intercept $b$ and the class-averaged radial correlation function $\overline{\overline{\rho}}(r)$, so that the 
decision trace provides the value of the decision function based only on contributions between 0 and $r$;  $\lim_{r\rightarrow\infty} d(r)$ gives the value which is ultimately used for classification.

PCovR models were constructed using the same feature data and decision values as the corresponding SVM model. The optimization target was the PCovR loss (sum of regression and projection losses) based on a three-component latent space.
The convex hull was constructed in the space defined by the
framework energies and the first two PCovR components for the PCovR model based on the full 6.0 \AA{} SOAP power spectrum feature vectors and the corresponding four-class decision functions for the classification exercise on these same feature vectors.

The train-test split of the databases was slightly modified based on the ultimate goal of the machine learning exercise (see details in the SI). To account for the imbalance in the class populations, 
we employed class-specific misclassification penalties for the SVM models: 
the penalty for a given class is weighted inversely proportional 
to the true class proportion in the train set. 
In contrast, class imbalance in the PCovR models 
was accounted for through replication of minority samples to achieve approximate class parity. This approach was preferred over undersampling the majority class, because the smallest number of minority class samples 
in a given training fold was very low, i. e. less than 20 structures. 

\newcommand{\noopsort}[1]{}

\end{document}